\documentclass[aps,prd,twocolumn,10pt,nofootinbib,showkeys,superscriptaddress]{revtex4-1}
\usepackage{amsmath}
\usepackage{amssymb}
\usepackage{graphicx}
\usepackage{color}
\usepackage{hyperref}
\usepackage{graphics}
\usepackage{epsfig}

\usepackage[english]{babel}
\usepackage[utf8]{inputenc} 

\newcommand{\nk}{\textbf{k}}

\newcommand{\dphi}{\delta \phi}
\newcommand{\x}{\textbf{x}}

\newcommand{\bra}{\langle}
\newcommand{\ket}{\rangle}

\newcommand{\RI}{\text{R,I}}

\newcommand{\barr}{\begin{eqnarray}}
\newcommand{\earr}{\end{eqnarray}}

\graphicspath{ {../} }

\begin{document}

\title{Emergent universe: tensor perturbations within the CSL framework}

\author{Octavio Palermo}

\affiliation{Grupo de Cosmolog\'{\i}a, Facultad
	de Ciencias Astron\'{o}micas y Geof\'{\i}sicas, Universidad Nacional de La
	Plata, Paseo del Bosque S/N 1900 La Plata, Argentina.}

\author{Martin Miguel Ocampo}

\affiliation{Grupo de Cosmolog\'{\i}a, Facultad
	de Ciencias Astron\'{o}micas y Geof\'{\i}sicas, Universidad Nacional de La
	Plata, Paseo del Bosque S/N 1900 La Plata, Argentina.}

\author{Gabriel R. Bengochea}
\email{gabriel@iafe.uba.ar} \affiliation{Instituto de Astronom\'\i
	a y F\'\i sica del Espacio (IAFE), CONICET - Universidad de Buenos Aires, (1428) Buenos Aires, Argentina}

\author{Gabriel Le\'{o}n}
\email{gleon@fcaglp.unlp.edu.ar }
\affiliation{Grupo de Cosmolog\'{\i}a, Facultad
	de Ciencias Astron\'{o}micas y Geof\'{\i}sicas, Universidad Nacional de La
	Plata, Paseo del Bosque S/N 1900 La Plata, Argentina.}
\affiliation{CONICET, Godoy Cruz 2290, 1425 Ciudad Aut\'onoma de Buenos Aires, Argentina. }

\begin{abstract}

We calculate the primordial power spectrum of tensor perturbations, within the emergent universe scenario, incorporating a version of the Continuous Spontaneous Localization (CSL) model as a mechanism capable of: breaking the initial symmetries of the system, generating the perturbations, and also achieving the quantum-to-classical transition of such perturbations. We analyze how the CSL model modifies the characteristics of the \emph{B}-mode CMB polarization power spectrum, and we explore their differences with current predictions from the standard concordance cosmological model. We have found that, regardless of the CSL mechanism, a confirmed detection of primordial \emph{B}-modes that fits to a high degree of precision the shape of the spectrum predicted from the concordance $\Lambda$CDM model, would rule out one of the distinguishing features of the emergent universe. Namely, achieving a best fit to the data consistent with the suppression observed in the low multipoles of the angular power spectrum of the temperature anisotropy of the CMB. On the contrary, a confirmed detection that accurately exhibits a suppression of the low multipoles in the \emph{B}-modes, would be a new feature that could be considered as a favorable evidence for the emergent scenario. In addition, we have been able to establish an upper bound on the collapse parameter of the specific CSL model used.

\end{abstract}

\maketitle

\section{Introduction}
\label{intro}

The indirect detection of primordial gravitational waves, through features imprinted in the cosmic microwave background (CMB) polarization (i.e. \emph{B}-modes), would be taken as an extraordinary experimental support for the inflationary model of the early universe. Although this weak signal has not yet been detected, many projects are already operating or have been proposed to measure the primordial \emph{B}-modes polarization of the CMB; and thanks to some of them, we already have valuable constraints on, for instance, the so-called tensor-to-scalar ratio parameter \cite{SPT19,POLARBEAR14,POLARBEAR15,ACTPol16,Henderson15,BICEP218,BICEP316,Simons18,LSPE20,Abazajian19,Hazumi19,PICO19,SPIDER21,QUBIC20a,QUBIC20b,Planck18b,BICEPKeck22,Campeti22,Belsunce22,Paoletti22}.

The current cosmological model provides us the possibility of being able to reconstruct the evolution of the universe, which includes a quantum description of the early universe, where during an inflationary phase the seeds of cosmic structure are generated, as a result of small quantum fluctuations of the fields in their vacuum state. These predictions have been verified with very high precision in analyses, for example, of the CMB \cite{Planck18b, Planck18c}. According to this scheme, our description of the early universe starts from a phase where both, spacetime and the quantum state of the fields, have symmetries such that they correspond to a perfectly isotropic and homogeneous situation. One might then ask: how is it that we ended up in a situation where small inhomogeneities appeared and the aforementioned symmetries were lost? This question is intimately linked to the so-called ``measurement problem'' in quantum physics, or more frequently referred to as the quantum-to-classical transition, see e.g. \cite{Wigner63,Omnes,Maudlin95,Becker,Norsen,Durr,Albert,okon14,barrett19,Sabine22}. Although we think that quantum theories give a more fundamental description of nature than classical theories, and therefore in reality such a transition would never occur, we want to find a mechanism that allows us to better understand how it is that under certain circumstances the classical description is an excellent approximation for our purposes. Such a proposal must keep in mind that observers and measuring apparatuses cannot be fundamental notions, in the search for a theoretical description of the early universe where neither existed \cite{Hartle93}. On the other hand, such a mechanism must also be able to account for how it is that the initial symmetries of the cosmological situation at hand were lost \cite{Bell81,PSS06,Sudarsky11}\footnote{A pedagogical review on this subject can be find in \cite{Bengochea20}.}.

In \cite{Maudlin95}, the author manages to approach the measurement problem in such a way that the different proposed alternatives fall into three claims mutually inconsistent: (A) The wave-function of a system is complete, i.e. it specifies all of the physical properties of a system, (B) the wave-function always evolves in accord with a linear dynamical equation, i.e. the Schr\"odinger equation, and (C) measurements always have determinate outcomes. The different ways of approaching the subject have been studied by many authors, for example through the addition of \emph{hidden variables} \cite{bohm,Valentini08b,Neto12,goldstein15,Neto18,Valentini19}, or works based on decoherence and/or ``many-worlds" interpretation of quantum mechanics \cite{kiefer09,halliwell,kiefer2,polarski,Everett,mukhanov2005}. In this work we will choose to address a possible solution to these problems within the framework of \emph{objective collapse theories}, modifications to the Schr\"odinger equation with the aim to alter the evolution of the wave function. These modifications negate claim (B), and the fundamental idea behind them is that the collapse of the wave-function would happen without the involvement of external agents, such as observers or measuring devices \cite{Pearle76,Ghirardi86,Pearle89,Diosi87,Diosi89,Penrose96}. In particular, we will use a version of the Continuous Spontaneous Localization (CSL) model \cite{Pearle76,Pearle89}. These models should still be taken as effective approximations and not as fundamental theories, since they are under development, but with great activity in recent years and have shown interesting and encouraging results \cite{Bassi1, Bassi2,Carlesso22,Daniel10,Sudarsky11,Tejedor12,Pedro13,Bengochea15,Leon15,Leon16,Leon17,Landau12,Susana13,Benetti16,Bengo17,Pedro18,Benito18,Picci19,Lucila15,Mariani16,Maj17,ModosB,Bouncing16,Josset17,Leon2020,MartinShadow,Bengo20Letter,Martin20R,Bassi21,Martin21,GLGB21}, showing that there is an extensive landscape of possibilities open \cite{Bengo20Long}.

Some proposals that aim to give a description of the early universe, with some modifications to the inflationary paradigm or with particular features, have been analyzed in some depth (e.g. \cite{Molina99,Kanekar2001,Khoury01,Steinh02,Batte04,Bojowald04,Biswas2005,Peter02,Batte15,Lilley15,Branden17,Matsui19,Barrau20,Agullo20,Agullo21,BrandenRev11,Branden18,Branden21}). Among them, one of the alternatives that seeks to escape the singularity theorems \cite{HawkingEllis1973,Wald84,Borde1993,Borde1994a,Borde1994b,Borde1996,Borde1997,Guth99, Borde2003}, and with some renewed motivation as a result of the recent debate about what is the spatial curvature of the universe \cite{Rasanen2014,Ratra17b,Ratra18b,Ratra18c,Ratra18d,Ratra19a,Ratra19b,Handley2019,Efstathiou19,Riess2019,Silk2020,Valentino2020,Efstathiou20,DiValentino2020a,Benisty20,Vagnozzi20A,Vagnozzi20B,Dhawan21}, is the known \emph{emergent universe} \cite{Ellis1}.

Built in the framework of General Relativity, the emergent universe (EU) model is one in which the dynamics is driven by a scalar field minimally coupled to gravity, but whose initial phase has been modified \cite{Ellis1, Ellis2}. A spatially closed universe starts from a static initial state with a finite size. At a certain time, the universe begins to evolve into a super-inflation phase, then slow-roll inflation occurs, and finally give rise to the standard hot-Big Bang. Different variants analyzing its viability and open issues have been studied in recent years, which has allowed to put into play an attractive possibility that deserves further exploration \cite{Ellis3,Gibbons1987,Barrow2003,Mukherjee05,Mukherjee06,Banerjee07,Boehmer07,Parisi2007,Campo1,Debnath08,Banerjee08,Beesham09,Wu2009,Campo2,Paul10,Paul2010,Zhang10,Paul2011,Chatto11,Campo3,Campo4,Labrana2012,Cai2012,Rudra12,Ghose12,Liu13,Aguirre13,Cai2014,Ataz14,Zhang14,Bag14,Huang15,Boehmer15,Labrana15,Kho16,Zhang16,Rios16,Khodadi16,Barrau18,Labrana19,Labrana21}.

A characteristic that was pointed out in \cite{Biswas13} is that a phase of super-inflation (i.e. a period where the Hubble parameter increases with time) prior to slow-roll inflation could be related to the suppression of power in the low CMB multipoles. In \cite{CSLEmerg21}, some of us showed that implementing the CSL mechanism to the emergent universe scenario introduces extra modifications in the CMB temperature angular spectrum. Specifically, the angular spectrum in the low multipoles sector can exhibit a suppression or an increment, something different from what happens in the case of the standard EU, where the super-inflation phase only causes the spectrum curve to decrease on large angular scales. Such a scenario gives good predictability to the CSL collapse proposal in the emergent universe model, distinguishing it from preceding works.

The work \cite{CSLEmerg21} was carried out within the framework of semiclassical gravity, where it is well known from previous works \cite{Lucila15,Maj17,ModosB} that tensor perturbations would be practically null, remarkably consistent with current observational constraints \cite{Planck18b,BICEP218,SPIDER21,BICEPKeck22,Campeti22,Belsunce22}. On the other hand, the exploration of studies in the framework of standard quantization (SQ), where a joint metric-matter quantization is performed, has also been done in many works, e.g. \cite{Martin12,Tejedor12B, Das13,Leon16,Mariani16,Bouncing16,MartinShadow,Bengo20Letter,Martin20R,Bassi21,Martin21}.


Let us mention some words about this particular point. Since we still do not fully understand the quantum nature of gravitation, it is interesting to study how our predictions depend on how one implements the theoretical ideas under different quantization approaches (i.e. semiclassical vs SQ).

In an earlier paper \cite{Bengo20Long}, some of us analyzed pros and cons of each of them and in particular we pointed out the problems facing the SQ approach. However, if one insists on following this approach, it is necessary to know what would be the effects of incorporating a collapse mechanism that allows solving the aforementioned problems related to the origin of the primordial inhomogeneities.

Under certain appropriate assumptions, previous works have explored situations that incorporated collapses under the SQ approach, e.g. \cite{Tejedor12B, Das13, Leon16, Bouncing16}, and it was found that in some cases the results were different from the standard inflationary model and in other cases the results were similar. Furthermore, despite the technical and conceptual difficulties presented by the SQ approach \cite{Bengo20Long}, one can find CSL models that agree with all the empirical data to date. However, we must emphasize once again that, even in the cases in which one obtains results similar to the standard approach without collapses, the last one has no physical process that clearly explains the following: how the primordial perturbations emerged, how the breaking of the initial symmetries (both of spacetime and of the initial vacuum state) occurred, and  how the so called ``quantum-to-classical transition'' of those perturbations took place. The latter of course refers to the passage from dealing with quantum fields to treat them as classical fields under a very good approximation.

On the other hand, within the SQ approach, let us note that there are different ways to incorporate collapses, e.g. \cite{Martin12, Das13,Das14}. In fact, in a previous paper \cite{Leon16} some of us showed that the results and predictions for primordial spectra could differ, depending on how the role of the CSL mechanism was implemented in the standard inflationary scenario. Those findings serve as the primary motivation for the present work.  Specifically, we wish to explore what is the prediction for the primordial tensor power spectrum employing the SQ scheme within the framework of the emergent universe. The CSL model considered here will be incorporated in a manner consistent with \cite{Leon16}. Another motivation is purely empirical, i.e. if \emph{B}-modes were to be detected observationally, and the semiclassical quantization approach faces some tension\footnote{The inflationary CSL model within the semiclassical gravity framework could face some tension if there is a confirmed detection of the primordial $B$ modes and is consistent with the standard prediction of slow roll inflation. This is because, in the former case, the predicted amplitude of the tensor power spectrum is of order $10^{-12} \epsilon^2$ \cite{Maj17,ModosB}, while the standard prediction is of order $10^{-9}\epsilon$, where $\epsilon$ is the slow roll parameter $\epsilon \ll 1$.}, then it is necessary to know the details of the differences (or similarities) in the theoretical predictions between the emergent universe with and without collapses incorporated, following the SQ scheme.

In summary, here we decided to extend our previous analysis of the emergent universe with the inclusion of the CSL model, to explore what would be the prediction for the primordial power spectrum associated to the tensor modes and the tensor-to-scalar ratio, within the framework of the joint quantization of metric and matter. In the present work we also set out to analyze if there are characteristics of the CSL that are manifested or not in the \emph{B}-mode polarization spectrum of the CMB, which can be distinguished from the standard $\Lambda$CDM cosmological model in the observations of future projects.

We divided this work as follows. We begin in section \ref{secdos} reviewing the basic concepts of the emergent universe, as well as the theory behind the CSL model. We also show the predicted primordial tensor power spectrum within this framework. Next, in section \ref{sectres} we show and discuss our results. Finally, in section \ref{conclusions}, we present our conclusions analyzing the main points that stand out to us. With respect to the conventions, we will use the $(-,+,+,+)$ signature for the spacetime metric and units such that $c=1=\hbar$ and $M_P^2 \equiv (8\pi G)^{-1}$.

\section{Emergent universe in the CSL framework revisited}
\label{secdos}

\subsection{A brief theoretical background}

In this subsection, we start reviewing the theoretical background of the CSL model and how it is applied to the case of tensor perturbations into the emergent universe (EU) framework.

We will be working under the same assumptions of \cite{Ellis1,CSLEmerg21}, i.e. the action of General Relativity with a scalar field $\phi$ minimally coupled to gravity and driving the early expansion. A typical scalar potential, as shown in \cite{Ellis2}, is $V(\phi)=(4\pi G)^{-1}(e^{C\phi}-1)^2$. In the reconstruction of such potential, the evolution of the background given by the scalar factor $a(t)\simeq a_0+A\:e^{H_0 t}$ was assumed, with $a_0 >0$ the (initial) radius of the Einstein static universe, $C$ and $A$ positive constants, and $H_0$ is the Hubble parameter at the onset of slow-roll inflation. In \cite{Rios16}, it was shown that the universe evolves from an Einstein static state to a (slow-rolling regime) de Sitter type of expansion. That is, the temporal evolution given by Friedmann and Klein-Gordon equations leads the system towards an attractor, where $H$ tends to a constant and $\dot{\phi}^2 \to 0$. The de Sitter type inflation is followed by a re-heating phase and finally the universe enters the standard expansion of the hot Big Bang.

A generic characteristic of the EU scenario is that, before to the slow-roll inflation, there is a phase of super-inflation where the Hubble parameter increases with time, i.e. $\dot{H}>0$. On the other hand, the spatial curvature is quickly negligible after a few e-foldings and furthermore slow-roll inflation can always be made to end for some negative value of $\phi$ \cite{Ellis2}. As in \cite{Labrana15,CSLEmerg21}, in this work we will make a first approach to the analysis and therefore we will neglect the contributions of the space curvature to the primordial perturbations\footnote{See, for instance, Appendix A of Ref. \cite{Labrana15} for details about this subject.}.

As usual in perturbation analysis, we will separate the metric and the scalar field into a homogeneous background plus small perturbations, i.e. $g_{\mu \nu} = g_{\mu \nu}^{(0)} + \delta g_{\mu \nu}$ and $\phi= \phi_0 + \dphi$. At first order in the tensor metric perturbations, the corresponding line element is
\begin{equation}\label{metricapert}
d s^{2}=a^{2}\left[-d \eta^{2}+\left(\delta_{i j}-h_{i j}\right) d x^{i} d x^{j}\right].
\end{equation}

In these coordinates, the scale factor results,
\begin{equation}\label{aeta}
a(\eta)=\frac{a_{0}}{1-e^{a_{0}H_{0}\eta}}.
\end{equation}
From here on, a prime over variables will denote derivative with respect to conformal time $\eta$.

Since the CSL theory is based on a stochastic non-linear modification of the Schr\"{o}dinger equation, it will be convenient to carry out the quantization in the Schr\"odinger picture. Therefore, the first step will be to write the total Hamiltonian of the system. As it is known, tensor perturbations represent gravitational waves, and they are characterized by a symmetric, transverse and traceless tensor field. These properties lead to the existence of only two degrees of freedom, i.e. two polarizations. But, as each polarization term is independent, and as each polarization leads to the same result, we will work with only one polarization. Then, we will just multiply by a factor of two the spectrum associated to an individual case, at the end of our calculations, to obtain the final result.

The action for tensor perturbations can be obtained by expanding the Einstein action up to the second order in transverse, traceless metric perturbations $h_{ij}(\mathbf{x},\eta)$   \cite{muk92,mukhanov2005}. Then, writing these perturbations in Fourier modes, $h_{ij}(\mathbf{k},\eta)=e_{ij}(\mathbf{k}) h_{\mathbf{k}}(\eta)$, where $e_{ij}(\mathbf{k})$ is a time-independent polarization tensor (which is  symmetric, traceless and transverse to $\nk$), and if we also perform the change of variable
\begin{equation}\label{MStensor}
h_{\mathbf{k}}(\eta)\equiv\frac{2}{M_{p}\left(e_{j}^i  e_{i}^j\right)^{1 / 2}} \frac{ v_{\mathbf{k}}(\eta)}{a(\eta)},
\end{equation}
the action up to the second order for these perturbations can be written as $S_{v}^{(2)}=\frac{1}{2} \int d \eta d^{3} \mathrm{k} \mathcal{L}_{v}$, where
\begin{equation}\label{lagrangiano}
\mathcal{L}_{v}= \left[v_{\mathbf{k}}^{\prime} v_{-\mathbf{k}}^{\prime}-\left(k^{2}-\frac{a^{\prime \prime}}{a}\right) v_{\mathbf{k}} v_{-\mathbf{k}}\right]
\end{equation}
Notice that, since $v(\mathbf{x},\eta)$ describes a real scalar field, we have that $v_{\mathbf{k}}^{*}=v_{\mathbf{-k}}$. On the other hand, the momentum canonical to $v_{\mathbf{k}}$ is $p_{\mathbf{k}}=\frac{\partial{\mathcal{L}_{v}}}{\partial{v_{\mathbf{k}}^{*'}}}$.

Therefore, the total Hamiltonian in Fourier space results
\begin{equation}
H=\int_{\mathbb{R}^{3+}} d^{3} k \quad\left[p_{\mathbf{k}}^{*} p_{\mathbf{k}}+v_{\mathbf{k}}^{*} v_{\mathbf{k}}\left(k^{2}-\frac{a^{\prime \prime}}{a}\right)\right]
\end{equation}

To work with real variables, it will be convenient to separate the canonical variables into their real and imaginary parts as:
\begin{equation}\label{descomposicion}
v_{\mathbf{k}} \equiv \frac{1}{\sqrt{2}}\left(v_{\mathbf{k}}^{\mathrm{R}}+i v_{\mathbf{k}}^{\mathrm{I}}\right), \quad p_{\mathbf{k}} \equiv \frac{1}{\sqrt{2}}\left(p_{\mathbf{k}}^{\mathrm{R}}+i p_{\mathbf{k}}^{\mathrm{I}}\right)
\end{equation}

Next, the fields $v_{\mathbf{k}}$ and $p_{\mathbf{k}}$ are promoted to quantum operators, satisfying the equal time commutator relation given by
\begin{equation}\label{conmutacion}
\left[\hat{v}_{\mathbf{k}}^{s}, \hat{p}_{\mathbf{k}^{\prime}}^{s'}\right]=i \delta\left(\mathbf{k}-\mathbf{k}^{\prime}\right)\delta_{ss'}
\end{equation}
where $s=R,I$ and $\delta_{ss'}$ is Kronecker's delta. Using \eqref{descomposicion} and \eqref{conmutacion}, the Hamiltonian results to be $\hat{H}=\int_{\mathbb{R}^{3+}}d^{3}k (\hat{H}_{\mathbf{k}}^{R}+\hat{H}_{\mathbf{k}}^{I})$, with
\begin{equation}\label{hamiltoniano}
    \hat{H}_{\mathbf{k}}^{R,I}=\frac{(\hat{p}_{\mathbf{k}}^{R,I})^{2}}{2}+\frac{(\hat{v}_{\mathbf{k}}^{R,I})^{2}}{2}\left(k^{2}-\frac{a^{\prime \prime}}{a}\right)
\end{equation}

In order to apply the CSL model into the EU scenario, we will follow the approach presented in \cite{Pedro13,Leon16} for the inflationary case. The temporal evolution characterizing each mode of the quantum field is given by:
\begin{equation}\label{cslevolution}
\begin{aligned}
\left|\Phi_{\mathbf{k}}^{\mathrm{R}, \mathrm{I}}, \eta\right\rangle &=\hat{T} \exp \left\{-\int_{\tau}^{\eta} d \eta^{\prime}\left[i \hat{H}_{\mathbf{k}}^{\mathrm{R}, \mathrm{I}}\right.\right.\\
&\left.\left.+\frac{1}{4\lambda_{k}}\left(\mathcal{W}_{\mathbf{k}}^{\mathrm{R}, \mathrm{I}}(\eta)-2 \lambda_{k} \hat{v}_{\mathbf{k}}^{\mathrm{R}, \mathrm{I}}\right)^{2}\right]\right\}\left|\Phi_{\mathbf{k}}^{\mathrm{R}, \mathrm{I}}, \tau\right\rangle
\end{aligned}
\end{equation}
where $\hat T$ is the time-ordering operator and $\tau$ denotes the conformal time at the beginning of the EU regime. This modification of the Schr\"{o}dinger equation allows it to be possible to attain a \emph{collapse} in the relevant operators corresponding to the Fourier components of the field. We will further assume linearity in the collapse generating operator, so that the CSL will act on each mode of the field independently. The stochastic field $\mathcal{W}_\nk(\eta)$ depends on the conformal time and $\nk$, so it could be regarded as a Fourier transform on a certain stochastic spacetime field $\mathcal{W}(\x,\eta)$. On the other hand, the second main CSL equation is the one that gives the probability for the stochastic field, i.e. the Probability Rule
\begin{equation}\label{cslprobabF}
	P(\mathcal{W}_{\nk}^{\RI}) d\mathcal{W}_{\nk}^{\RI} =   \bra \Phi_{\nk}^{\RI} , \eta | \Phi_{\nk}^{\RI}, \eta \ket \prod_{\eta'=\tau}^{\eta-d\eta} \frac{ d \mathcal{W}_{\nk}(\eta')^{\RI}}{\sqrt{2 \pi \lambda_k/d\eta}}.
\end{equation}

From the CSL evolution given by Eq. \eqref{cslevolution}, it can be seen that we have chosen the field variable $\hat{v}_\nk^\RI$ as the collapse generating operator. Operationally, what happens is that the evolution given by the CSL mechanism drives the initial state of the system to an eigenstate of $\hat{v}_\nk^\RI$, with a certain collapse rate given by the CSL parameter $\lambda_{k}$. As usual in the framework of a joint metric-matter quantization of the perturbations, we will adopt the point of view that the classical characterization of $h_{\mathbf{k}}$ is an adequate description if the quantum state is sharply peaked around some particular value. As a consequence, the classical value corresponds to the expectation value of $\hat{h}_{\mathbf{k}}$ \cite{Tejedor12B}. More precisely, the CSL collapse mechanism will lead to a final state $|\Psi \ket$ such that the relation
\begin{equation}\label{igualdadchingonah}
h_{\mathbf{k}} =  \bra \Psi| \hat{h}_{\mathbf{k}} |  \Psi \ket
\end{equation}
is valid.  It is evident that a quantization of $v_{\nk}$ from the action built with Eq. \eqref{lagrangiano} yields a quantization of $\hat h_{\nk}$. In other words, equations \eqref{MStensor} and \eqref{igualdadchingonah} imply that:
\begin{equation}\label{relacion hk vk}
h_{\mathbf{k}}(\eta)=\frac{2}{M_{P}\left(e_{j}^i  e_{i}^j\right)^{1 / 2}} \frac{\langle \hat{v}_{\mathbf{k}}(\eta)\rangle}{a(\eta)}
\end{equation}
Thus, Eq. \eqref{relacion hk vk} relates the quantum field variable $\hat v_{\nk}$ to the amplitude of the classical tensor mode $h_{\nk}$. In particular, Eq. \eqref{relacion hk vk}  serves as a justification for choosing $\hat v_\nk$ as the collapse operator. Let us note here that when the quantum state is the vacuum, we have that $\bra 0|\hat{v}_{\mathbf{k}} |0\ket=0$ and then the tensor perturbation is $h_{\mathbf{k}}=0$ (the same occurs to the primordial curvature perturbation, namely to the scalar perturbations of the metric). It is only after the state has evolved, according to the CSL mechanism, that generically $\bra  \hat{v}_{\mathbf{k}} \ket \neq 0$ and the tensor perturbation is generated (as well as the scalar curvature perturbation).  The quantum expectation value $\bra \hat v_{\nk} \ket$ acts as a source for the tensor perturbation. This illustrates how the self-induced collapse provided by the CSL model can generate the primordial perturbations and achieve the quantum-to-classical transition.

In Fourier space, the wave functional $\Phi[v,\eta]$ can be factorized into mode components $\Phi[v_{\mathbf{k}},\eta]=\prod_{\mathbf{k}} \Phi_{\mathbf{k}}^R[v_{\mathbf{k}}^{R},\eta] \times \Phi_{\mathbf{k}}^I[v_{\mathbf{k}}^{I},\eta]$. On the other hand, since the ground state of the Hamiltonian \eqref{hamiltoniano} is a Gaussian, and because the Hamiltonian and the CSL evolution equation are quadratic in both $\hat{v}_{\nk}^{\textrm{R,I}}$ and $\hat{p}_{\nk}^{\textrm{R,I}}$, the wave functional at any time can be written in the form:
\begin{equation}\label{funcional}
\Phi^\RI [v_{\nk}^\RI,\eta] =
	\exp[-A_k(\eta) (v_{\nk}^\RI )^2 + B_k^\RI(\eta) v_{\nk}^\RI + C_k^\RI (\eta) ]
\end{equation}
with initial conditions given by
\begin{equation}\label{condiciones iniciales}
A_k(\tau) = \frac{k}{2}, \qquad B_k^\RI(\tau)=0, \qquad C_k^\RI (\tau)=0,
\end{equation}
corresponding to choose as the initial state of the field the standard Bunch-Davies (BD) vacuum.

\subsection{Primordial tensor power spectrum}

In this subsection, we will focus on deriving a prediction for the primordial spectrum of the tensor perturbations. Notice that, since the equations have the same mathematical structure, we will closely follow the steps shown in \cite{CSLEmerg21} for the scalar spectrum case.

The tensor power spectrum associated to $h_{ij}(\mathbf{k},\eta)$ is defined as
\begin{equation}\label{defps}
\overline{h_{j}^{i}(\mathbf{k}) h_{i}^{j *}(\mathbf{k'})}\equiv\frac{2 \pi^{2}}{k^3} P_{t}(k) \delta(\mathbf{k}-\mathbf{k}^{\prime})
\end{equation}
where $P_{t}(k)$ is the dimensionless power spectrum and the bar appearing in the last equation means an ensemble average over possible realizations of the stochastic field $h_{ij}(\mathbf{k},\eta)$. It should be remembered here that, in the CSL framework, each realization is associated to a particular realization of the stochastic process characterizing the collapse.

By using Eq. \eqref{relacion hk vk} we arrive at
\begin{equation}\label{aux1}
\overline{h_{j}^{i}(\mathbf{k}) h_{i}^{j *}(\mathbf{k'})}= \frac {4}{M_{P}^{2} a^{2}} E(\mathbf{k},\mathbf{k'}) \overline{\langle \hat{v}_{\mathbf{k}}\rangle \langle \hat{v}_{\mathbf{k'}}\rangle ^{*}}
\end{equation}
where $E(\mathbf{k},\mathbf{k'})$ is a scalar factor dependent on the polarization tensor, defined by
\begin{equation}
    E(\mathbf{k},\mathbf{k'})=\frac{e^{i}_{j}(\mathbf{k})e^{j}_{i}(\mathbf{k'})^{*}}{\left(e^{m}_{n}(\mathbf{k})e^{n}_{m}(\mathbf{k})\right)^{\frac{1}{2}} \left[\left(e^{r}_{s}(\mathbf{k'})e^{s}_{r}(\mathbf{k'})\right)^{\frac{1}{2}}\right]^{*}}
\end{equation}
that satisfies $E(\mathbf{k},\mathbf{k})=1$.

From definition \eqref{defps} and Eq. \eqref{aux1}, we can identify an equivalent tensor power spectrum as:
\begin{equation}\label{aux2}
P_{t}(k) \delta(\mathbf{k}-\mathbf{k}^{\prime})=\frac{2 k^{3}}{\pi^{2} M_{P}^{2} a^{2}} E(\mathbf{k},\mathbf{k'}) \overline{\langle \hat{v}_{\mathbf{k}}\rangle \langle \hat{v}_{\mathbf{k'}}\rangle ^{*}}
\end{equation}

Taking into account the real and imaginary parts of $\hat v_\nk$, the ensemble average in \eqref{aux2} is
\begin{equation}\label{aux3}
\overline{\langle \hat{v}_{\mathbf{k}}\rangle \langle \hat{v}_{\mathbf{k'}}\rangle ^{*}}= \frac{1}{2}(\overline{\langle \hat{v}_{\mathbf{k}}^{\mathrm{R}}\rangle ^{2}} + \overline{\langle \hat{v}_{\mathbf{k}}^{\mathrm{I}}\rangle ^{2}} ) \delta(\mathbf{k}-\mathbf{k}^{\prime})
\end{equation}

Since $\overline{\langle \hat{v}_{\mathbf{k}}^{R}\rangle^{2}}=\overline{\langle \hat{v}_{\mathbf{k}}^{I}\rangle^{2}}$, we will omit the indexes R,I from now on. On the other hand, using the main equations of the CSL model, Eqs. \eqref{cslevolution} and \eqref{cslprobabF}, one obtains:
\begin{equation}\label{aux4}
    \overline{\langle\hat{v}_{\mathbf{k}}\rangle^{2}}=\overline{\langle \hat{v}_{\mathbf{k}}^{2}\rangle}-\frac{1}{4 \operatorname{Re}(A_{k})}
\end{equation}

Then, substituting Eqs. \eqref{aux3} and \eqref{aux4} into Eq. \eqref{aux2}, we find that the power spectrum can be expressed as:
\begin{equation}\label{masterPS2}
P_{t}(k)=\frac{2k^{3}}{\pi^{2} M_{P}^{2} a^{2}}\left(\overline{\langle \hat{v}_{\mathbf{k}}^{2}\rangle}-\frac{1}{4 \operatorname{Re}(A_{k})} \right)
\end{equation}

The final steps of the calculation consist of explicitly calculating the two terms on the right of the Eq. \eqref{aux4}. Since this calculation is similar to the one performed in \cite{CSLEmerg21}, we refer the reader to Appendix \ref{appA} where we have reproduced the details particularized for the present case. The final result for the tensor power spectrum turns out to be:
\begin{equation}\label{psMain}
    P_{t}(k)= A_t \chi^2 \left|F(\chi)\right|^2 C(k),
\end{equation}
where $\chi \equiv k/a_0 H_0$, and we have defined the amplitude of the tensor power spectrum as
\begin{equation}\label{At}
A_t\equiv\frac{2 H_{0}^{2}}{\pi^{2} M_{P}^{2}}.
\end{equation}
The functions $F(\chi)$ and $C(k)$ are defined in Appendix \ref{appA}.

\section{Results: impact on the $B$-modes angular spectrum}
\label{sectres}

In this section, we shall proceed to examine possible observational features in the \emph{B}-modes of the CMB polarization spectrum, as a consequence of the introduction of the CSL mechanism in the emergent universe model.

First, notice that the mathematical structure of the tensor power spectrum $P_t(k)$ of Eq. \eqref{psMain} is similar (except for the amplitude) to the scalar spectrum $P_s(k)$ shown in Eq. (36) of \cite{CSLEmerg21}. Therefore, we will proceed to do the analysis in a similar manner; and in particular, under the same assumptions discussed in the aforementioned work. In this way, the primordial tensor power spectrum, in order to include the small scale dependence normally associated with the tensor spectral index, can be expressed as:
\begin{equation}\label{PPSexacta}
P_t (k) = A_t   \chi^2 |F(\chi)|^2 C(k)  \left( \frac{k}{k_P} \right)^{n_t}
\end{equation}
where $k_P$ is a pivot scale, which we set as $k_P = 0.05$ Mpc$^{-1}$, and $n_t$ is the tensor spectral index. As shown in Appendix \ref{appA}, the last expression can be approximated by Eq. \eqref{PsAprox}. Then,
\begin{equation}\label{TPS}
P_{t}(k)  \simeq A_t \frac{\chi^2}{(1+\chi)^2 }  \frac{\lambda_k |\tau|}{k}\left( \frac{k}{k_P} \right)^{n_t}.
\end{equation}

A well known result is that, using the action for the perturbations $h_{ij}$ and the action of the scalar field $v$ (corresponding to the Mukhanov-Sasaki field variable), together with the definitions of the tensor and scalar (curvature) power spectra, leads to $r=16 \epsilon$, with $\epsilon$ the first slow-roll parameter of inflation. As a consistency check, we can see that from the amplitude of the scalar spectrum found in \cite{CSLEmerg21} (i.e. $A_s=H_0^2/8\pi^2 \epsilon M_P^2$, together with the amplitude of the tensor power spectrum found in this work, $A_t=2H_0^2/\pi^2 M_P^2)$, yields $r\equiv A_t/A_s=16\epsilon$ accordingly.

Second, we will also assume the same parameterization for the collapse rate $\lambda_k$ as in \cite{CSLEmerg21} (see that Ref. for the motivation of such a choice), i.e
\begin{equation}
	\label{eq:param_lambdak}
	\lambda_k = \lambda_0 \, ( k + B)
\end{equation}
where $\lambda_0 = 1.029$ Mpc$^{-1}$, this numerical value is motivated by the fact that such a value is within the range allowed by current laboratory experiments \cite{sandro2017}. Also, the parameterization of the form \eqref{eq:param_lambdak}, i.e. at linear order in $k$,  is necessary to achieve the approximation \eqref{TPS}, specifically in the function $C(k)$. For the initial conformal time, we have chosen $|\tau| \simeq 10^8$ Mpc; in this manner, the condition $k |\tau| \gg 1$ is fulfilled for the modes $k$ within the range of observable interest: $10^{-6}$ Mpc$^{-1}$ $\leq k \leq 1$ Mpc$^{-1}$.  On the other hand, we fix $a_0H_0 = 2 \times 10^{-4}$ Mpc$^{-1}$ and the parameter $B\geq 0$ will take values between $10^{-3}$ and $10^{-4}$ Mpc$^{-1}$, which represent the preferred values obtained from the analysis in \cite{CSLEmerg21}. In particular, those values yield theoretical curves of the scalar power spectrum $P_s(k)$ and the CMB temperature angular power spectrum that seem to be consistent with the latest data from \textit{Planck} collaboration \cite{Planck18a} (see Figs. 1 and 2 of \cite{CSLEmerg21}). Given that in the present section we are seeking to perform the complementary analysis using the tensor modes, the choice of these values for the parameters ensures that what was found for the scalar case remains valid and consistent. We remind the reader that $B=0$ corresponds to practically ``turning off'' the effects of the collapse mechanism\footnote{It should be remembered that, in our case, collapses through the CSL model are always present; since, strictly speaking, non-collapse implies that $\lambda_k=0$.}. In that case, the tensor power spectrum obtained would correspond to the one from the original EU model presented in \cite{Labrana15}, which we will name the \textit{original emergent universe} model (OEU). In this way, $B$ quantifies small deviations from the OEU reflecting the inclusion of the CSL model. Also, we include in each Figure the {\em canonical model}, which will be used as a second reference. The canonical model corresponds to the standard $\Lambda$CDM cosmological model, with parameters coming from the latest data from \textit{Planck} collaboration \cite{Planck18c}. At the $68\%$ confidence level these values are: $\Omega_bh^2 = 0.02236$, $\Omega_{c}h^2 = 0.1202$, $H_{\rm today} = 67.27 \,{\rm km \, s^{-1} \, Mpc^{-1}}$, $A_s = 2.101 \,\times\, 10^{-9}$ and the optical depth $\tau_d = 0.0544$. For the tensor power spectrum, we have $A_t=r A_s$, with the tensor-to-scalar ratio parameter $r = 0.036$ at 95\% confidence \cite{BICEPKeck22}. The latter implies that, by the consistency relation $n_t = -r/8$, we can use $n_t = -0.0045$.

As was mentioned in the Introduction, an important feature of the emergent universe is that a phase of super-inflation prior to slow-roll inflation could be related to the suppression of power in the low CMB multipoles. In \cite{CSLEmerg21}, some of us showed that implementing the CSL collapse proposal to the emergent universe scenario (through the parameter $B$) introduces extra modifications in the CMB temperature angular spectrum. Specifically, the angular spectrum in the low multipoles sector ($l < 50$) can exhibit a suppression or an increment, a different feature from what is generically produced in the emergent universe, which only decreases the curve spectrum at large angular scales.

Our next step is to analyze if there are characteristics of the CSL mechanism that are manifested or not in the \emph{B}-modes angular spectrum of the CMB, which can be distinguished from the standard $\Lambda$CDM cosmological model in the observations of future projects. To achieve this, we modify the public Code for Anisotropies in the Microwave Background (CAMB) software \cite{Lewis:1999bs}.

\begin{figure}[h]
	\centering
	\includegraphics[width=0.45\textwidth]{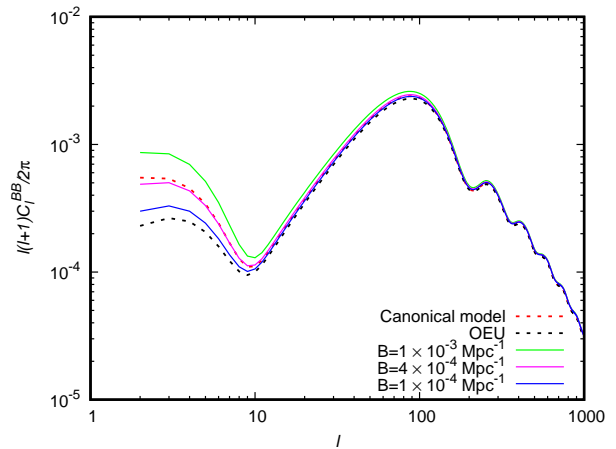}
	\caption{Predicted \emph{B}-mode spectra for different values of the collapse parameter $B$ of the CSL model implemented in the emergent universe scenario. The \emph{canonical} $\Lambda$CDM model and the \emph{original emergent universe} (OEU) are also shown.}
	\label{fig_1}
\end{figure}

Figure \ref{fig_1} depicts the resulting \emph{B}-modes CMB polarization power spectrum, for different values of the collapse parameter $B$. There, it can be seen that by varying $B$ there is an excess or suppression of the angular spectrum for low multipoles $l < 10$, which is precisely the region of the $C_l^{BB}$ spectrum where primordial gravitational waves contribute the most. In the case of the $C_l^{TT}$ spectrum, exactly the same behavior occurred (Fig. 2 of \cite{CSLEmerg21}), but the fact that the curve was above or below the \emph{canonical model} did not allow one to rule out any of these possibilities. At most, from the known fact of 'anomalies' at low multipoles \cite{copi5, Perivola21}, one could say that the set of parameter values of the model that suppress such low multipoles have some observational advantage. However, if we now also take into account the $C_l^{BB}$ plot, we see that from a certain value  $B_{max} \simeq 4 \times 10^{-4}$,  the emergent universe + CSL curve passes above the canonical one. That is, \emph{values higher than $B_{max}$ would already be ruled out because they exceed the canonical spectrum}. Let us recall that the canonical $C_l^{BB}$ spectrum was constructed using the maximum observationally allowed constraint on the tensor-to-scalar ratio $r$. Thus, with the current constraints on the primordial \emph{B}-modes, we can jointly use the $C_l^{TT}$ and $C_l^{BB}$ spectra to further constrain the $B$ parameter of the CSL collapse model.

\begin{figure}[h]
\centering
\includegraphics[width=0.45\textwidth]{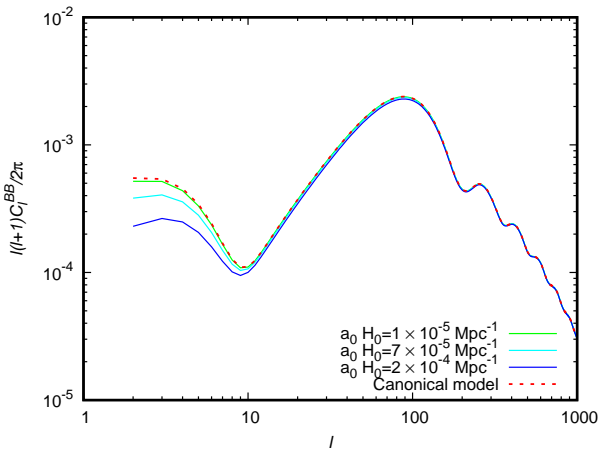}
\caption{Predicted \emph{B}-mode spectra, when the CSL effect is (practically) turned off (i. e. $B=0$) for different values of the emergent model parameter $a_0H_0$. The \emph{canonical} $\Lambda$CDM model is also displayed.}
\label{fig_2}
\end{figure}

Another interesting case to analyze is what would happen if there was a confirmed detection of primordial \emph{B}-modes and  the standard theoretical $C_l^{BB}$ curve, i.e. the one obtained from the canonical model, fits the data to a high degree of precision. To analyze this case, we will turn off the collapse effect by setting $B=0$, so that the collapse effect is practically eliminated from the emergent universe model case (also we can always arrange $|\tau|$, $\lambda_0$ and  $H_0$ so that the amplitude $A_t$ remains unchanged).

In Fig. \ref{fig_2}, we see that the value $a_0 H_0 = 2 \times 10^{-4}$ Mpc$^{-1}$ (typically assumed for the analyses) would be ruled out under the mentioned hypothesis. However, it is possible to decrease the value of $a_0 H_0$ in such a way that the $C_l^{BB}$ curve of the OEU model fully approaches the canonical one. If this happens, the OEU prediction for the $C_l^{BB}$ would be indistinguishable from the canonical one and consistent with the data. The remarkable aspect about this effect is that, decreasing the value of $a_0 H_0$, also causes the 'tail' of the low multipoles corresponding to the temperature spectrum (i.e. the $C_l^{TT}$) to increase, approaching the canonical one. In other words, \emph{a confirmed detection of primordial B-modes that matches accurately the canonical $C_l^{BB}$ curve, would rule out a main feature of the OEU model, namely that it can solve the problem associated with a lack of power at large angular scales observed in the $C_l^{TT}$ spectrum}.

On the contrary, if a confirmed detection of the primordial \emph{B}-modes shows a suppression of low multipoles with respect to the canonical $C_l^{BB}$ curve, as it apparently does for the $C_l^{TT}$ one, then we find that the OEU model could be preferred by the data over the canonical one, precisely because it has the characteristic that it suppresses the low multipoles simultaneously in both spectra. Finally, notice that 'turning on' the parameter $B$ of the CSL model, would not have any effect that substantially changes the previous analysis. That is, if $B>0$, then it would not be possible to suppress the low multipoles in the $C_l^{TT}$ spectrum and, at the same time, not affect the $C_l^{BB}$ in the same way.

\section{Conclusions}
\label{conclusions}

In this work, we have studied the primordial tensor power spectrum in the emergent universe scenario, incorporating a particular version of the CSL model as a mechanism capable of generating and explaining the quantum-to-classical transition of the primordial perturbations; that is, to achieve a regime in which quantum quantities can be described to a sufficient accuracy by their classical counterparts.

The search and detection of the CMB \emph{B}-modes is an active field currently involving many collaborations. We have shown that non-trivial features might be detectable in the \emph{B}-modes CMB polarization power spectrum within the emergent universe scenario, either with or without the additional effects of the CSL model considered here.

By varying the collapse parameter $B$, there is an excess or suppression of the angular spectrum for low multipoles $l < 10$, which is precisely the region of the $C_l^{BB}$ spectrum where primordial gravitational waves contribute the most. This result is similar to the case of the $C_l^{TT}$ spectrum analysed in \cite{CSLEmerg21}. However, in the tensor case, we see that above a certain value given by $B_{max} \simeq 4 \times 10^{-4}$, the curve corresponding to the emergent universe with the addition of the CSL model, passes above the canonical one. That is, \emph{values higher than $B_{max}$ would already be ruled out by present data, because they produce curves that exceed the canonical tensor spectrum, which was constructed using the current maximum constraints allowed for the tensor-to-scalar ratio $r$}. This result confirms that, using the spectra $C_l^{TT}$ and $C_l^{BB}$ together enables to establish further observational constraints on the parameter $B$ of this version of the CSL collapse model.

Another main result is that, regardless of the CSL mechanism, a confirmed detection of primordial \emph{B}-modes that fits to a high degree of precision the shape of the spectrum predicted from the concordance $\Lambda$CDM model, would rule out one of the distinguishing features of the emergent universe. Namely, producing a best fit to the data consistent with the observed suppression in the low multipoles of the angular power spectrum of the temperature anisotropy of the CMB. Although the emergent universe model would not be ruled out, the values allowed for $a_0H_0$ would not be those that grants the advantage to the emergent universe over the $\Lambda$CDM model; specifically, to achieve a better fit in the low multipole region of the $C_l^{TT}$ spectrum. In fact, in that case, the $\Lambda$CDM and the emergent universe predicted  spectra  (i.e. the $C_l^{TT}$ and $C_l^{BB}$ theoretical curves) would be indistinguishable. On the contrary, only for a confirmed detection of the primordial \emph{B}-modes that shows a suppression of low multipoles with respect to the canonical $C_l^{BB}$ spectrum, the emergent universe model could be favored by the data.

To conclude, let us note that as long as precise data of the $C_l^{BB}$ spectrum are not available, from the mere fact of the decrease in the power of the low-$\ell$ tensor spectrum, one cannot conclusively say whether a collapse mechanism was at play in the early universe or not, because such an effect is achieved in the emergent universe with or without the CSL model. If, in the event that, in addition to a suppression in the low multipoles, some different feature in the shape of the tensor spectrum is detected, then we would be able to distinguish in a more precise manner  between the cases $B=0$, i.e. the OEU model,  and the case with non-zero $B$ collapse parameter. However, we must emphasize that even in the case in which the standard OEU and the one with a CSL model are not distinguishable, the OEU lacks the mechanism that allows explaining the breaking of symmetries and the quantum-to-classical transition. Furthermore, we must remember that it is in conjunction with observations of the scalar temperature spectrum that we will be able to achieve a distinction between models. As mentioned in \cite{CSLEmerg21}, we expect that in the scalar case the parameter $B$ is not centered at $B=0$, which would help distinguish our proposal from both the canonical $\Lambda$CDM case and the one explored in \cite{Labrana15}.

\appendix

\section{Annex calculations of the power spectrum}\label{appA}

We show here, guided by the analysis shown previously in \cite{CSLEmerg21}, the intermediate steps of the calculation to arrive at the final expression in Eq. \eqref{psMain} of the tensor power spectrum, and its approximate version given by Eq. \eqref{TPS}.

Let us start with the second term of Eq. \eqref{aux4}. The quantity $[\textrm{Re}(A_{k})]^{-1}$ represents the variance of the field variable, which in turn is related to the width of the wave functional \eqref{funcional}. From Eq. \eqref{cslevolution} and the wave functional \eqref{funcional}, one can obtain an equation of evolution for this quantity, which results
\begin{equation}
A_k'=-2 i A_k^2+\frac{i}{2}\left(k^{2}-\frac{a^{\prime \prime}}{a}\right)+\lambda_k
\end{equation}
It is convenient to rewrite this last equation, making the change of variables given by $A_k \equiv f'/(2i f)$. In this way, we have
\begin{equation}\label{ecuacionf}
 f'' + \left(q^2 - \frac{a''}{a} \right) f = 0
\end{equation}
with:
\begin{equation}\label{defq}
	q^2 \equiv k^2 \left(1 - 2i \frac{ \lambda_k}{k^2}\right).
\end{equation}
A solution to Eq. \eqref{ecuacionf} with the Bunch-Davies initial conditions given by Eq. \eqref{condiciones iniciales} can be found, which results
\begin{equation}\label{solf}
f = \frac{e^{-i q \eta}}{\sqrt{2k} (1- e^{a_0 H_0 \eta})  } \: \:_2F_1  (q_-, q_+,b; e^{a_0 H_0 \eta}),
\end{equation}
where $\:_2F_1$ is the hypergeometric function, $q\pm$ and $b$ defined by
\begin{equation}\label{qpm}
q\pm\equiv-1-\frac{iq}{a_{0} H_{0}}\pm\sqrt{1-\left(\frac{q}{a_{0}H_{0}}\right)^{2}}
\end{equation}
\begin{equation}\label{be}
b\equiv1-\frac{2iq}{a_{0} H_{0}}
\end{equation}

Now, returning to the original variable $A_{k}$, we find that $[\textrm{Re}(A_{k})]^{-1}=(\lambda \eta)^{-1}$. On the other hand, by virtue of the definition for $f$ results,
\begin{equation}
\textrm{Re} [A_k(\eta)] = \frac{W}{|f|^2 4 i}
\end{equation}
being $W=f^{'}f^{*}-f^{'*}f$ the corresponding Wronskian. Notice that if $\lambda_k = 0$, then $W = i$ for all $\eta$, and $q = k$.

Next, we will focus on the first term of Eq. \eqref{aux4}. Here, it will be convenient to define the following quantities:
\begin{equation}
Q \equiv \overline{\langle\hat{v}_{\mathbf{k}}^2\rangle}, \quad R \equiv \overline{\langle\hat{p}_{\mathbf{k}}^2\rangle}, \quad S \equiv \overline{\left\langle\hat{p}_{\mathbf{k}} \hat{v}_{\mathbf{k}}+\hat{v}_{\mathbf{k}} \hat{p}_{\mathbf{k}}\right\rangle}
\end{equation}
Then, from the CSL equations we can obtain equations for $Q, R$ and $S$, which result:
\begin{equation}
\begin{aligned}
    &Q^{\prime}=S\\
    &R^{\prime}=-w_{k}(\eta) S+\lambda_{k}\\
    &S^{\prime}=2R-2Q w_{k}(\eta)
\end{aligned}
\end{equation}
where we name $w_k (\eta) \equiv k^2 - a''/a$. This is a linear system of coupled differential equations. The general solution will be a particular solution to the system plus a solution to the homogeneous equation ($\lambda_k=0$). The solution results:
\begin{equation}\label{solgeneralQ}
Q(\eta)=C_{1} v_{1}^{2}+C_{2} v_{2}^{2}+C_{3} v_{1} v_{2}+Q_{p}
\end{equation}
where $C_{1}$, $C_{2}$ and $C_{3}$ are found by imposing the initial conditions corresponding to the Bunch-Davies vacuum state: $Q(\tau)=1/2k,
R(\tau) = k/2$, and $S(\tau)=0$. On the other hand, the functions $v_{1}$ and $v_{2}$ are two linearly independent solutions of $v''+w_{k}v=0$, and $Q_{p}$ is a particular solution of
\begin{equation}\label{ecQp}
Q_{p}^{\prime \prime \prime}+4 w_{k} Q_{p}^{\prime}+2 w_{k}^{\prime} Q_{p}=2 \lambda_{k}
\end{equation}

The exact solutions $v_{1}$ and $v_{2}$ are:
\begin{equation}
\begin{aligned}
v_{1}(\eta)&=\frac{e^{-i k \eta}}{\sqrt{2 k}\left(1-e^{a_{0} H_{0} \eta}\right)}{ }_{2} F_{1}\left(k_{-}, k_{+}, b ; e^{a_{0} H_{0} \eta}\right)\\
v_{2}(\eta)&=v_{1}(\eta)^{*}
\end{aligned}
\end{equation}
with $k_\pm$ and $b$ are defined in the same manner as in \eqref{qpm} an \eqref{be} but replacing $q \to k$.

We should note here that an exact solution to Eq. \eqref{ecQp} is difficult to find. However, given the regimes of interest in the present work (the initial static regime where de BD conditions are imposed, and the de Sitter phase where de power spectrum is evaluated), we can find approximate solutions. In the static regime $w_{k}\simeq k^{2}$, and in the de Sitter one $w_{k}=k^{2}-2/\eta^{2}$. In these two regimes mentioned, $Q_p$ can be approximated by
\begin{equation}
Q_{p}(\eta)\simeq\frac{\lambda_{k} \eta}{2k^{2}}
\end{equation}

With this in hand, the constants that appear in Eq. \eqref{solgeneralQ} can be calculated, which turn out to be:
\begin{equation}
C_{1}=\frac{-i \lambda_{k}}{4 k^{2}} e^{2 i k \tau}, \quad C_{2}=C_{1}^{*}, \quad C_{3}=1-\frac{\lambda_{k} \tau}{k}.
\end{equation}

With all this, we can now obtain the power spectrum \eqref{masterPS2}. Before that, let us note that if $\lambda_k = 0$ then $P_{t}(k) =0$, because $Q(\eta) = (4 \textrm{Re} [A_k(\eta)])^{-1}$ exactly in that case. This result is consistent with our point of view in which, if there is no collapse, then the metric perturbations are zero.

Then, by considering the modes in the super-Hubble limit ($-k\eta \to 0$), the power spectrum \eqref{masterPS2} can be written as
\begin{equation}\label{ps}
    P_{t}(k)=\frac{2 H_{0}^{2}}{\pi^{2} M_{P}^{2}} \chi^2 \left|F(\chi)\right|^2 C(k),
\end{equation}
with:
\begin{equation}
    \chi\equiv\frac{k}{a_{0} H_{0}}
\end{equation}
\begin{equation}\label{Fposta}
F(\chi)\equiv\frac{2\Gamma\left(1-2 i \chi\right)}{\Gamma\left(2-i \chi-\sqrt{1-\chi^{2}}\right) \Gamma\left(2-i \chi+\sqrt{1-\chi^{2}}\right)}
\end{equation}
\begin{equation}\label{Cposta}
    C(k)\equiv 1+\frac{\lambda_k}{k} |\tau|+\frac{1}{2} \sin(2\delta) \frac{\lambda_k}{k^2}
\end{equation}

\begin{equation}
    \delta\equiv\arctan\left(\frac{\textrm{Im} F}{\textrm{Re} F}\right)-\chi a_{0} H_{0} |\tau|
\end{equation}

To arrive at equation \eqref{ps} we have multiplied by two due to the different polarizations, and approximated the scale factor by $a \approx - \frac{1}{\eta H_{0}}$.


As discussed in depth in \cite{CSLEmerg21}, numerical calculations set a restriction for implementing the exact equation \eqref{ps}. However, for the whole $k$ range of observational interest, namely
\begin{equation}\label{rangok}
	10^{-6} \: \textrm{Mpc}^{-1}\leq k \leq 1 \: \textrm{Mpc}^{-1},
\end{equation}
 one can use an approximate expression. Let us see how to implement it.

Starting from Eq. \eqref{Fposta}, we can rewrite this function as follows:
\begin{equation}
     F(\chi)= \frac{2\Gamma(x_1)}{\Gamma(x_2)\Gamma(x_3)}
\end{equation}
with:
\begin{equation}
    x_1=1-2i\chi,
\end{equation}
\begin{equation}
    x_2=2-i\chi-\sqrt{1-\chi^2},
\end{equation}
\begin{equation}
    x_3=2-i\chi+\sqrt{1-\chi^2}.
\end{equation}

In Eq. \eqref{ps} we have $|F(\chi)|^2$, so given the properties of the Gamma function, we explicitly write it as:
\begin{equation}
    |F(\chi)|^2 = \frac{4|\Gamma(x_1)|^2}{|\Gamma(x_2)|^2|\Gamma(x_3)|^2}.
\end{equation}

We now consider the two asymptotic regimes for $\chi$, i.e $\chi^2\gg1$ and $\chi^2\ll1$.

\begin{itemize}
    \item If we consider $\chi^2\gg1$, then $\sqrt{1-\chi^2} \approx i\chi$. Therefore, $x_2 \approx x_1+1$ and $x_3 \approx 2$. Using these approximations in the exact expression of $|F(\chi)|^2$ \eqref{Fposta}, and taking into account that $\Gamma(z+1)=z\Gamma(z)$ and $\Gamma(2)=1$, we obtain
\begin{equation}
    |F(\chi)|^2 \approx \frac{4|\Gamma(x_1)|^2}{|x_1|^2|\Gamma(x_1)|^2} =\frac{4}{1+4\chi^2} \approx \frac{1}{\chi^2}.
\end{equation}
    \item On the other hand, in the regime $\chi^2\ll1$, we can approximate $\sqrt{1-\chi^2}\approx 1-\chi^2/2$; hence $x_2\approx 1-i\chi$ and $x_3\approx 3-i\chi=x_2+2$. By using the property $|\Gamma(1+bi)|^2= {\pi b}/{\sinh(\pi b)}$, we can write
\barr
|F(\chi)|^2&\approx&\frac{4|\Gamma(x_1)|^2}{|x_2+1|^2|x_2|^2|\Gamma(x_2)|^4}\\
&=&\frac{4\tanh(\pi \chi)}{\pi \chi (4+\chi^2)(1+\chi^2)} \approx 1 + O(\chi^2)
\earr
where at the end we have performed a Taylor series around $\chi=0$.
\end{itemize}

\begin{figure}[h]
\centering
\includegraphics[width=0.45\textwidth]{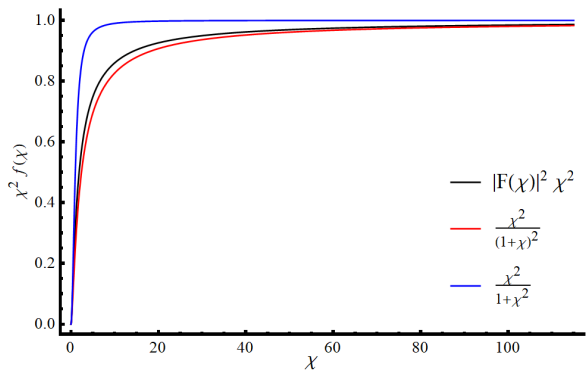}
\caption{Two possible approximations for the function $\chi^2|F(\chi)|^2$ (black line) mentioned in the text. Option (i) corresponds to $f(\chi) = g(\chi)$ (blue line) and option (ii) corresponds to $f(\chi) = h(\chi)$ (red line). We observe that $\chi^2 h(\chi)$ is a better approximation to the exact solution in the full regime.}
\label{fig_3}
\end{figure}

Thus, we have two asymptotic regimes: (i) $|F(\chi)|^2 \to 1/\chi^2$ for $\chi^2 \gg 1$ and (ii) $|F(\chi)|^2 \to 1$  for $\chi^2 \ll 1$. In order to match smoothly these two regimes, we can consider two options: $g(\chi)={1}/{(1+\chi^2)}$ and $h(\chi)={1}/{(1+\chi)^2}$. Plotting both functions, together with the exact form $|F(\chi)|^2$, we can see that $h(\chi)$ is a better approximation and thus we will use that
\begin{equation}\label{Fapprox}
	\chi^2|F(\chi)|^2\approx \frac{\chi^2}{(1+\chi)^2},
\end{equation}
in the power spectrum (see Fig. \ref{fig_3}).

On the other hand, $C(k)$ is defined by Eq. \eqref{Cposta}, which we can rewrite it as
\begin{equation}\label{Cposta2}
    C(k)=1+\frac{\lambda_{k} |\tau|}{k}  \left(1 +\frac{\sin(2\delta) }{2k |\tau|} \right).
\end{equation}
Taking into account that as an initial condition in the Bunch-Davies vacuum we have $k|\tau|\gg1$ (and that also $\left|\frac{\sin{2\delta}}{2}\right|\le \frac{1}{2}$), we can neglect the last term, thus,
\begin{equation}\label{Capprox1}
	C(k) \approx 1+\frac{\lambda_{k} |\tau|}{k}.
\end{equation}
Next, we recall that the parameterization used for $\lambda_{k}$ is given by $\lambda_{k}=\lambda_{0}(k+B)$. Note that if $B=0$, the predicted angular spectrum is indistinguishable from the one corresponding to the OEU model. Therefore, we can think of the parameter $B$ as representing a small deviation the  OEU model due to the inclusion of the collapses. Using the aforementioned parametrization in Eq. \eqref{Capprox1}, we have
\begin{equation}\label{Capprox2}
	C(k) \approx  \lambda_0 |\tau| \left(1 + \frac{B}{k} + \frac{1}{\lambda_0 |\tau|}   \right).
\end{equation}
In the analysis presented in Sec. \ref{sectres}, the value of the parameter $B$ is within the interval $[10^{-4},10^{-3}]$ Mpc$^{-1}$, while $k$ is within the range of observational interest $[10^{-6},1]$ Mpc$^{-1}$. Consequently,  $B/k$ is bounded within $[10^{3},10^{-4}]$.  Moreover, since we have chosen $\lambda_0 \approx 1$  Mpc$^{-1}$ and $|\tau| \approx 10^8$ Mpc, then  $1/\lambda_0 |\tau| \approx 10^{-8}$.  Thus,
\begin{equation}\label{Capprox3}
	C(k) \approx  \lambda_0 |\tau| \left(1 + \frac{B}{k} \right) =\frac{\lambda_{k}}{k} |\tau|.
\end{equation}

Finally, using  approximations \eqref{Fapprox} and \eqref{Capprox3}, the tensor power spectrum \eqref{ps} becomes:
\begin{equation}\label{PsAprox}
    P_{t}(k)\approx \frac{2 H_{0}^{2}}{\pi^{2} M_{P}^{2}}\frac{\chi^2}{(1+\chi)^2}\frac{\lambda_{k}|\tau|}{k}.
\end{equation}


\begin{acknowledgements}
We thank the anonymous referee for his valuable suggestions and Mar\'ia  P\'ia Piccirilli for her useful aid in the numerical aspects of the work. G.R.B. is supported by CONICET (Argentina) and he acknowledges support from grant PIP 112-2017-0100220CO of CONICET (Argentina). G.L. is supported by CONICET (Argentina), and he also acknowledges support from the following project grants: Universidad Nacional de La Plata I+D G175 and PIP 112-2020-0100729CO of CONICET (Argentina).

\end{acknowledgements}

\bibliography{bibliografia}
\bibliographystyle{apsrev}
\end{document}